# CO Oxidation Facilitated by Robust Surface States on Au-Covered Topological Insulators

Hua Chen[1], Wenguang Zhu[1,2], Di Xiao[2], and Zhenyu Zhang[3,1]

[1]*Department of Physics and Astronomy,*

*The University of Tennessee, Knoxville, TN 37996, USA*

[2]*Materials Science and Technology Division,*

*Oak Ridge National Laboratory, Oak Ridge, TN 37831, USA*

[3]*ICQD/HFNL,*

*University of Science and Technology of China, Hefei, Anhui, 230026, China*

Surface states refer to electronic states emerging as a solid material terminates at a surface[1,2], and can be present in many systems, e.g., the free-electron like states on the (111) surfaces of noble metals[3], and the dangling bond derived states on semiconductor surfaces[4]. Despite their spatial proximity to material surfaces, surface states have been largely overlooked in fundamental understanding of surface catalysis and potential real-world applications, because of their vulnerability to local impurities or defects. In contrast, the recently discovered three-dimensional topological insulators (3DTI)[5,6,7,8] have exceptionally robust metallic surface states that are topologically protected against surface contamination and imperfection[9,10]. The robust topological surface state(s) (TSS) provides a perfect platform for exploiting novel physical phenomena and potential applications of surface states in less stringent environments. Here we employ first-principles density functional theory to demonstrate that the TSS can play a vital and elegant role in facilitating surface reactions by serving as an effective electron bath. We use CO oxidation on gold-covered $Bi_2Se_3$[11,12] as a prototype example[13,14,15,16], and first show that the TSS is preserved when a stable ultrathin Au film is deposited onto a Bi-terminated $Bi_2Se_3$ substrate. Furthermore, the TSS can significantly enhance the adsorption energy of both CO and $O_2$ molecules, by promoting different directions of electron transfer. For CO, the TSS accepts electrons from the CO-Au system, thereby decreasing the undesirable occupation of the CO antibonding states. For $O_2$, the TSS donates the needed electrons to promote the molecule towards dissociative adsorption. The present study adds a new arena to the technological potentials of 3DTI, and the central concept of *TSS as an electron bath* as revealed here may lead to new design principles beyond the conventional *d*-band theory of heterogeneous catalysis[17,18].

Aside from their spatial proximity, the *sp*- or *d*-band derived surface states can be located at or near the Fermi level ($E_F$) in energy space; consequently they can significantly influence the physical and chemical processes happening at the surfaces[19,20,21]. However, since they arise as a result of the different bonding environment at the surface from the bulk[1], normal surface states are easily destroyed by local modifications at the surfaces, e.g., presence of impurities, surface defects, surface reconstruction, or a change in the surface termination or orientation. This fundamental limitation has prohibited systematic studies of the potential role of surface states in surface reactions and catalysis, especially in more realistic environments.

In contrast, the exotic metallic TSS of the recently discovered 3DTI[5,6,7,8] are exceptionally robust compared to conventional surface states. The TSS arise from the nontrivial topology of the electron bands of the 3DTI, and their persistence is protected by time-reversal symmetry. Therefore, the TSS are insensitive to the structural details of the surface, and will persist as long as the bulk band gap embodying the TSS remains open. The robust TSS thus provides a perfect platform for investigating the catalytic role of surface states in less constrained environments.

In our study, we focus on CO oxidation by supported Au films, a prototype system of fundamental and practical importance in heterogeneous catalysis[13,14,15,16]. The use of the 3DTI $Bi_2Se_3$ [11,12] as the substrate, instead of conventional oxides such as $TiO_2$ or MgO, is to introduce additional surface electron states (the TSS) at $E_F$. As shown recently for the case of metal-induced semiconductor surface reconstruction[22], such extra states may serve as an electron bath to significantly modify the bonding configurations on the surface through proper electron transfer.

Crystalline $Bi_2Se_3$ has rhombohedral structure and its unit cell is composed of 3 weakly coupled quintuple layers (QL), each of which comprises of 5 alternating Bi and Se layers in a sequence Se-Bi-Se-Bi-Se[11,12]. The naturally cleaved surface is therefore the Se surface. In Figure 1a we plot the band structure around $E_F$ of a naturally cleaved 3 QL $Bi_2Se_3$ film. The surface states in the bulk band gap (~0.3 eV) have close to linear dispersion near $E_F$. This single "Dirac-cone" like band structure is a distinctive feature of the TSS of $Bi_2Se_3$ [11,12]. The small gap (~50 meV) opened at the Dirac point is due to the coupling between the two degenerate TSS on the two surfaces of a very thin 3DTI film[23] (also see Supplementary Section S1).

We use the Bi-terminated surface of $Bi_2Se_3$, to which Au binds more strongly than the naturally cleaved Se surface (Supplementary Section S2), to support Au atoms without formation of 3D Au clusters[15]. Experimentally it is possible to intentionally introduce a large amount of Se vacancies on the surfaces of $Bi_2Se_3$ or eventually form a complete Bi-terminated surface, due to the large vapor pressure difference between Bi and Se[23]. We therefore can use the ideal Bi-terminated surface in the present proof-of-principle study. The band structure of the Bi-terminated film (see Methods) is shown in Figure 1b. The two TSS Dirac bands still robustly persist and shift below $E_F$, agreeing with experiments[9,11]. Additionally, the degeneracy of the two Dirac bands is lifted, of which the upper and lower bands correspond to the TSS at the upper (Bi-terminated) and lower (Se-terminated) surface, respectively.

We choose 2 monolayers (ML, see Methods for definition) of Au deposited on the Bi surface of the $Bi_2Se_3$ film (Figure 1c inset) as a model system because of its optimal stability (see Supplementary Sections S2 and S3) for subsequent calculations. Note that the 2 ML of Au has roughly the same areal density as a single (111) layer of bulk Au. Figures 1c and 1d show the band structures of the Au-covered $Bi_2Se_3$ film without and with spin-orbit coupling (SOC), respectively. Two TSS Dirac bands emerge only when the SOC is switched on, confirming that the TSS indeed originates from the SOC of the bulk states[5,6]. This observation allows us to conveniently isolate the effects of the TSS by comparative studies with and without SOC. The shape of the two TSS bands near the Γ point closely resembles that of the TSS in Figure 1b for the Bi-terminated $Bi_2Se_3$ film despite the slight shift in their relative positions in energy. Therefore the TSS survives even if the $Bi_2Se_3$ surface is completely buried under the 2 ML Au film.

The calculated CO binding energies on the model system with and without SOC are listed in Table I. With SOC, the binding energy is considerably enhanced by 0.2 eV compared to that without SOC, accompanied by a decrease of the C-Au bond length from 2.029 Å to 1.981 Å. The enhanced CO binding with SOC is due to the electron transfer facilitated by the TSS. To see this effect we first compare the local density of states (LDOS) on the C atom of an adsorbed CO with and without SOC, shown in Figure 2a. The antibonding $2\pi^*$ states shift to higher energies with SOC, indicating decreased electron occupation (also see Supplementary Figure 4 for the spatial electron density difference plots), and hence enhanced CO-Au binding[18]. On the other hand, from Figure 2c, the top Dirac band, corresponding to the TSS on the Au-deposited Bi-terminated surface, shifts to lower energy after the adsorption of CO, indicating increased electron occupation. Taken together, net electrons are transferred to the TSS serving as an electron bath when CO is adsorbed on the surface.

The enhanced CO binding with SOC is not due to Au, though Au does have large SOC. To see this we consider two comparative cases: One is fixing the 2 ML Au film in vacuum with the $Bi_2Se_3$ substrate removed; the other is the unreconstructed Au(111) surface. From Table I, the CO binding energy differences with and without SOC are about one order of magnitude smaller (~0.03 eV) in these two latter cases. We also note that with SOC the CO binding energy on the Au-$Bi_2Se_3$ system is much larger than that on Au(111). Based on these results, we conclude that the TSS of the underlying $Bi_2Se_3$ enhance the CO binding by accepting the electrons that otherwise would have been transferred to the antibonding states of the CO-Au system.

Next we show that the TSS as an electron bath can also enhance the adsorption of $O_2$, but by invoking a different direction of electron transfer. On the Au-$Bi_2Se_3$ substrate, $O_2$ binding energy increases by 0.16 eV with SOC, which is also much larger than that on the freestanding Au film or Au(111) surface (Table I). The LDOS on one O atom of $O_2$ is shown in Figure 2b. The two groups of peaks below and above $E_F$ correspond to the spin-up and spin-down antibonding $2\pi^*$ states, respectively[24]. As the half-filled $2\pi^*$ states hybridize with the Au $d$ states, more electrons will be transferred to the $2\pi^*$ states and promote $O_2$ towards dissociation[24,25]. At the same time, the spin splitting of the $2\pi^*$ states will decrease due to the weakened O-O bond. In Figure 2b, both groups of the spin-splitted peaks shift toward $E_F$ after turning on SOC, indicating decreased spin splitting in the $O_2$ orbitals. Meanwhile, the O-O bond length increases from 1.289 Å without SOC to 1.299 Å with SOC. The increased electron occupation of the $2\pi^*$ states upon switching on SOC is not easily visible from Figure 2b, but is confirmed by the calculated increase in the relative spectral weight of the $2\pi^*$ DOS below $E_F$, equal to 0.56 with SOC and 0.55 without SOC. This difference is roughly equal to 0.04 $e$ per $O_2$ molecule, originated from the TSS (also see the electron density difference plots in Supplementary Figure 4). On the other hand, from Figure 2d, the top TSS Dirac band shifts upward compared to that without $O_2$ adsorption, indicating that electrons are transferred out of the TSS. Therefore, rather than accepting electrons as in the case of CO, the TSS now donates electrons and promotes $O_2$ towards dissociative adsorption on Au. Moreover, the adsorption energy of $O_2$ is now comparable to that of CO with a moderate strength, which is a desirable feature for easier reaction and high catalytic activity.

In studying adsorption of molecules on transition metal surfaces, the prevailing theoretical framework has been the $d$-band theory, according to which the chemical activity of a metal substrate is correlated with the position of its $d$-band center ($E_d$) in the energy spectrum[17,18]. Specifically the closer $E_d$ of the metal substrate to $E_F$, the stronger molecular adsorbates bind to the substrate. A major contribution to the energy gain as $E_d$ shifts up towards $E_F$ is from the decreased filling of the adsorbate's antibonding states as they are "pushed away" from $E_F$ by the $d$ bands. To see whether our results can be explained independently by the $d$-band theory, we have calculated $E_d$ of Au in the top layer of the 2 ML Au film (Figures 3a and 3b). We find that, in the present case, $E_d$ actually shifts down away from $E_F$ after switching on SOC, and the shift is even larger than that of the freestanding 2 ML Au film. These findings unambiguously rule out the $d$-band theory to be mainly operative in the present systems.

An implicit assumption of the $d$-band theory is that the $sp$ states around $E_F$ do not differ much among different systems. However, our present study offers a striking and unique counter example to this assumption. In particular, since these TSS are gapless and not completely filled, they can readily donate or accept electrons in order to lower the total energy upon molecular adsorption (see illustrations in Figures 3c-e). Such *sp*-band derived states near $E_F$, either due to extrinsic effects such as the 3DTI substrate studied here or associated with the transition metals themselves, can play a prominent role in surface reactivity, especially for late transition metals with deeper *d* bands. Furthermore, the delocalized *sp* surface states may also help to smooth the energy profile of the surface and lower the diffusion and reaction barriers of the adsorbed molecules[19,20], an intriguing aspect worth future investigations. The present study indicates manipulation of the *sp* surface states may be complementary to the guiding principles from *d*-band theory in searching and designing new catalysts, and 3DTI materials may offer rich opportunities for this purpose.

**METHODS**

Our density functional calculations are carried out using the Vienna ab initio simulation package (VASP)[26] with PAW potentials[27,28] and the generalized gradient approximation (PBE-GGA)[29] for exchange-correlation functional. The lattice constants of $Bi_2Se_3$ are adopted from experiments. The generic $Bi_2Se_3$ substrate is modeled by a slab of 15 atomic layers (3 QL), and a Bi-terminated substrate is realized by removing one outermost Se layer. The vacuum layers are over 20 Å thick to ensure decoupling between neighboring slabs. During relaxation, atoms in the lower 11 atomic layers are fixed in their respective bulk positions, and all the other atoms are allowed to relax until the forces on them are smaller than 0.01 eV/Å. A 7×7×1 k-point mesh is used for the 1×1 surface unit cell, and 3×3×1 for the 2×2 surface unit cell[30]. Binding energies are calculated as $\Delta E = E_{adsorbate+substrate} - E_{adsorbate} - E_{substrate}$, and with SOC unless otherwise stated. One monolayer (1 ML) corresponds to the same number of atoms as that in each atomic layer of $Bi_2Se_3$, which is equal to 0.48 times the atom density in a (111) layer of bulk Au. $E_d$ is calculated by $E_d = \int_{-\infty}^{\infty} PDOS_d(E) \times (E - E_F) dE / \int_{-\infty}^{\infty} PDOS_d(E) dE$, where $PDOS_d$ is the PDOS of Au $d$ bands.

**Supplementary Information** is available.

**Acknowledgements**

The authors thank Shunfang Li for helpful discussions. This work was supported by US National Science Foundation (Grant number 0906025), National Natural Science Foundation of China (Grant number 11034006), the Division of Materials Science and Engineering, Office of Basic Energy Sciences, US Department of Energy, and the UT/ORNL Joint Institute for Advanced Materials (JIAM fellowship). The calculations were performed at National Energy Research Scientific Computing Center of US Department of Energy.

**Author Information**

The authors declare no competing financial interests. Correspondence and requests for materials should be addressed to Zhenyu Zhang (zzhang1@utk.edu).

**FIGURE LEGENDS**

**Figure 1 | Band structures of bare and Au-covered $Bi_2Se_3$ films. a**, for a 3 QL $Bi_2Se_3$ film, where the shaded area is the bulk band structure projected to the 2D Brillouin zone. The inset shows the shape of the 2D Brillouin zone for different surface unit cells (Black solid lines—1×1; blue solid lines—2×2). In **a**, **b** and **d** the TSS are highlighted by the transparent blue lines. **b**, for a Bi-terminated $Bi_2Se_3$ film. **c** and **d**, for 2 ML Au deposited on the Bi-terminated surface without and with SOC, respectively. The inset in **c** shows the top and side views of the structure (only top 4 atomic layers of $Bi_2Se_3$ are shown). Yellow balls—Au; blue balls—Bi; dark blue balls—Se.

**Figure 2 | Electron transfer between adsorbed CO or $O_2$ and the TSS serving as an electron bath. a** and **b**, LDOS on the C atom of CO, and one O atom of $O_2$, featuring the energy range corresponding to the $2\pi^*$ states of CO and $O_2$, respectively. **c** and **d**, band structures of the CO and $O_2$ adsorbed 2 ML Au-$Bi_2Se_3$ film, shown in a reduced Brillouin zone corresponding to the 2×2 surface unit cell (Figure 1a inset). The TSS bands are highlighted by the transparent blue lines. The blue dot-dash lines indicate the position of the upper Dirac point in Figure 1d. The upper and lower panels in **e** and **f** are the top and side views of the atomic structures. Red balls—O; gray balls—C; yellow balls—Au; blue balls—Bi; dark blue balls—Se.

**Figure 3 | The TSS as an electron bath versus the *d*-band picture. a** and **b**, Density of states projected to the *d* orbital (PDOS) of an Au atom in the top layer of the 2 ML Au film with and without $Bi_2Se_3$ substrate, respectively**.** In **b** the Au film is fixed in space. The red and black dot-dash lines indicate the positions of $E_d$ with and without SOC,

respectively, which are -2.89 eV and -2.68 eV in **a**, and -2.69 eV and -2.57 eV in **b**. **c-e**, illustration of the role of the TSS in molecular adsorption. Blue—valence and conduction bands of the 3DTI support; green—antibonding states of adsorbed molecules; red—TSS; dark green—metal *d* bands. **c**, without the TSS, the support will not be involved in the electron transfer. **d** and **e**, with the TSS, electrons of the molecule-metal system can be transferred either to or from the partially filled metallic surface states, depending on which way can lower the total energy.

**TABLE**

**TABLE I. Binding energies (in eV) of CO and $O_2$ on three comparative substrates.** (1) 2 ML Au on Bi-terminated $Bi_2Se_3$, (2) freestanding 2 ML Au fixed in the same geometry as in (1), and (3) (unreconstructed) Au(111) surface. The Au(111) surface is modeled by a 4 layer slab, each layer comprising of 16 Au atoms, with the lower 2 layers fixed. The $CO/O_2$ coverage is 1/4 ML in (1) and (2), and 1/16 ML in (3). Definition of ML is in Methods. The geometries of adsorbed CO and $O_2$ are shown in Figures 2e and 2f, respectively.

|  |  | (1) Au-$Bi_2Se_3$ | (2) 2 ML Au | (3) Au(111) |
|---|---|---|---|---|
| CO | SOC | 0.49 | 0.95 | 0.30 |
|  | No SOC | 0.29 | 0.92 | 0.27 |
| $O_2$ | SOC | 0.23 | 0.52 | <0.01 |
|  | No SOC | 0.07 | 0.49 | <0.01 |

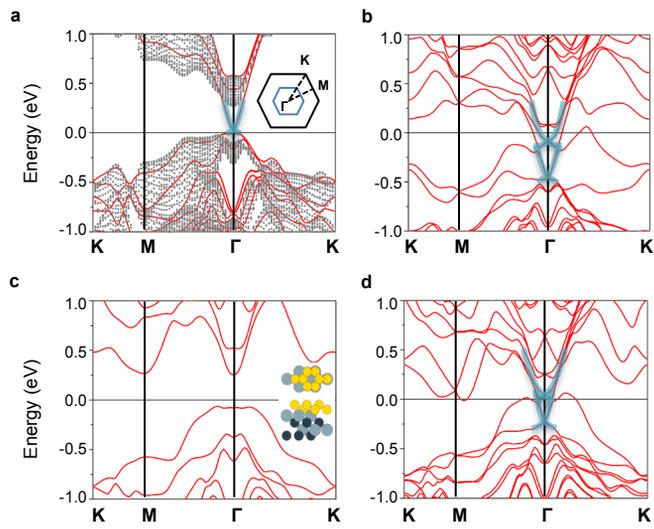

Fig. 1. Chen et al.

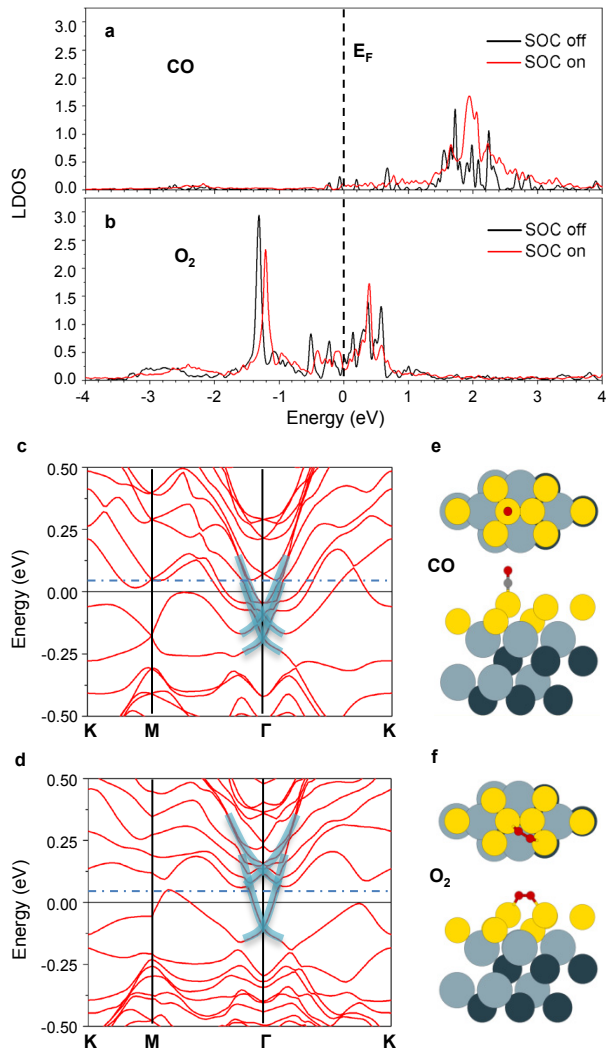

Fig. 2. Chen et al.

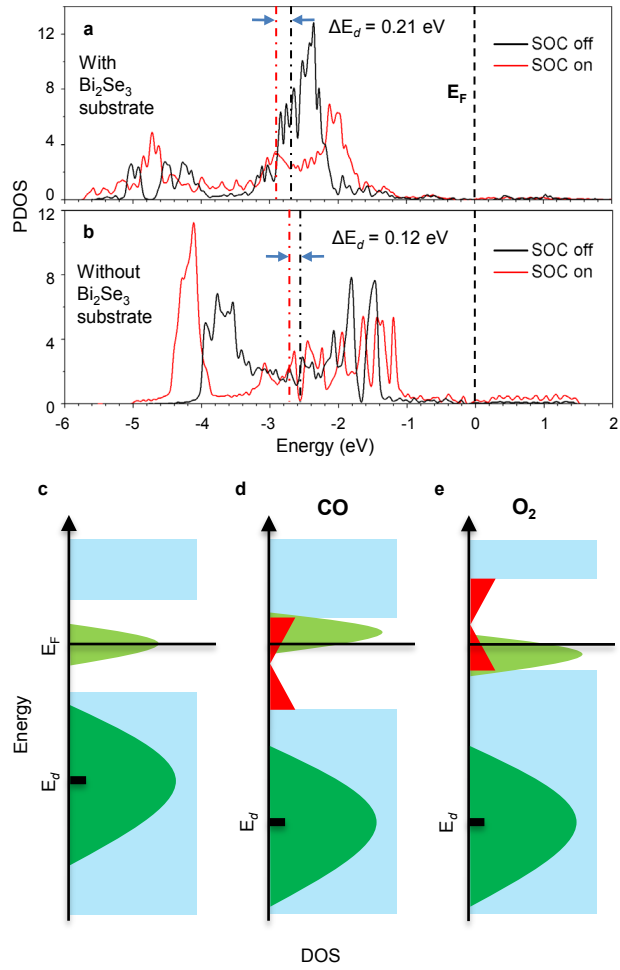

Fig. 3. Chen et al.